\documentclass[onecolumn,sort&compress,numbers]{els-mrw} 

\usepackage{amsmath,amssymb,amsfonts,amsthm,makeidx,graphicx}
\usepackage{txfonts}
\usepackage{helvet}

\begin{document}


\chapter{Non-particle dark matter}\label{chap1}

\author[1]{Anne M. Green}%

\address[1]{\orgname{University of Nottingham}, \orgdiv{School of Physics and Astronomy}, \orgaddress{University Park, Nottingham, NG7 2RD, UK}}


\maketitle

\begin{abstract}[Abstract]
We provide a pedagogical introduction to non-particle dark matter, focused on primordial black holes (PBHs), black holes that may form in the early Universe from large overdensities. First, we outline the key properties of PBHs and how they meet the requirements to be a dark matter candidate. We then overview how PBHs can form, in particular from the collapse of large density perturbations generated by inflation (a proposed period of accelerated expansion in the early Universe). Next, we describe how PBHs can be probed by observations. Finally, we conclude with a summary focused on the key open questions in the field.
\end{abstract}

\begin{keywords}
 	primordial black holes \sep dark matter \sep early universe cosmology
  
\end{keywords}


\begin{glossary}[Nomenclature]
	\begin{tabular}{@{}lp{34pc}@{}}
        CMB & Cosmic Microwave Background \\
        CO & Compact Object \\
        DM & Dark Matter\\
        MC & Magellanic Clouds \\
        MW & Milky Way \\
		(P)BH & (Primordial) Black Hole\\
        USR & Ultra-slow roll \\
	\end{tabular}
\end{glossary}

\section*{Objectives}
To understand that:
\begin{itemize}
	\item Primordial Black Holes (PBHs), black holes that may form from large overdensities in the early Universe, are a dark matter candidate.
	\item Large overdensities can be generated by inflation (a period of accelerated expansion postulated to have occurred shortly after the Big Bang), and also via various other mechanisms. 
	\item The abundance of PBHs of different masses can be probed by a variety of observations. Light PBHs are constrained by limits on the products of their Hawking evaporation, while planetary and Solar mass PBHs are constrained by stellar microlensing and other observations.
\end{itemize}

\section{Introduction}\label{sec-intro}

A wide range of astronomical and cosmological observations indicate that $85\%$ of the matter in the Universe is in the form of cold (non-relativistic), non-baryonic (not made of standard model particles) dark matter (DM). Traditionally new elementary particles, such as Weakly Interacting Massive Particles and axions, have been the most popular dark matter candidates. However DM might not be a new particle. Another possibility is Primordial Black Holes (PBHs), black holes formed in the early Universe, in the first second after the Big Bang. According to General Relativity, a black hole (a region where gravity is so strong that light can not escape) will form if a region containing mass, $M$, is smaller than its Schwarzschild radius, $R_{\rm s} = 2 G M/c^2$. As we will see in Sec.~\ref{sec-form}, in the early Universe there are various mechanisms that can generate sufficiently overdense regions. PBHs can have a very wide range of masses, unlike astrophysical black holes (formed at the end of the lifetime of massive stars or via the collapse of gas)
which are heavier than the Sun.

PBHs satisfy the key requirements of a viable DM candidate, namely they are cold, non-baryonic and stable on cosmological timescales. They are sufficiently massive that they are non-relativistic. Because they form before nucleons are produced at the quark-hadron phase transition, they are non-baryonic. While light PBHs evaporate and lose mass due to Hawking radiation (see Sec.~\ref{sec-evap}), PBHs with initial mass $M \gtrsim 10^{15} \, {\rm g}$ have lifetime longer than the age of the Universe~\cite{Hawking:1975vcx}. Unlike most other dark matter candidates, PBHs are not a new elementary particle. However, as we will see below, their formation does typically require  `Beyond the Standard Model' physics. 

PBHs were first studied in the late 1960s and early 1970s by Zeldovich and Novikov~\cite{Zeldovich:1967lct} and Hawking~\cite{Hawking:1971ei} and it was realised already in the 1970s that they were a viable DM candidate~\cite{Hawking:1971ei,Chapline:1975ojl}. The announcement in 2016 of the detection by LIGO-Virgo of gravitational waves from mergers of tens of Solar mass BHs~\cite{LIGOScientific:2016aoc} generated a surge of interest in PBHs as DM. It is now thought that these BHs are mostly, and potentially entirely, astrophysical BHs, see e.g.~Ref.~\cite{Franciolini:2021tla}. However, shortly after the announcement, several papers suggested that the BHs could be primordial and also make up some of the dark matter~\cite{Bird:2016dcv,Clesse:2016vqa,Sasaki:2016jop}. This sparked a wave of interest in PBHs in general and PBHs as DM in particular. See Ref.~\cite{Carr:2024nlv} for a detailed history of PBHs.

This chapter provides an introduction to Primordial Black Hole dark matter. In Sec.~\ref{sec-form} we explore the mechanisms via which PBHs can form. In particular, we focus on the collapse of large density fluctuations generated by inflation, a period of accelerated expansion postulated to have occurred in the very early Universe. In Sec.~\ref{sec-obs} we look at observational probes of the abundance of PBHs, before concluding with a summary and discussion of the key open questions in the field in Sec.~\ref{sec-conclusions}. Throughout we set $c=1$. As this is an Encyclopedia article, rather than citing an overwhelmingly large number of papers, instead we refer mainly to review papers and a few key papers in order to provide an entry point to the field for new researchers.

\section{Formation of Primordial Black Holes} \label{sec-form}

In this Section we explore the formation of PBHs in the early Universe, focusing on the most popular mechanism, the collapse of large density perturbations during radiation domination. 
In Sec.~\ref{sec-criteria} we look at the criteria an over-density must satisfy to form a PBH and the properties (in particular mass) of the resulting PBH.  In Sec.~\ref{sec-abundance} we calculate the initial and present-day abundances of PBHs, before briefly reviewing inflation models which generate sufficiently large density perturbations to form an interesting abundance~\footnote{For most practical purposes, an `interesting abundance' is a non-negligible fraction of the DM density.} of PBHs in Sec.~\ref{sec-inflation}. Finally, we conclude by briefly overviewing other PBH formation mechanisms in Sec.~\ref{sec-othermech}. 

\subsection{Criteria and properties}
\label{sec-criteria}

When the universe is radiation-dominated a region will collapse to form a PBH if it is sufficiently over-dense at horizon entry. Horizon entry occurs when a length scale becomes less than the Hubble scale, the distance light can have travelled since the Big Bang. Cosmologists use comoving coordinates, which scale out the expansion of the universe by dividing by the scale factor, $a$, 
and parameterise length scales in terms of the comoving wavenumber $k$, which is the inverse of the comoving length scale~\footnote{Further explanation can be found in Cosmology textbooks e.g.~Refs.~\cite{Baumann:2022mni,Huterer:2023mmv}.}. The distance light can have travelled since the Big Bang is of order $H^{-1}$, where $H=( {\rm d} a/{\rm d} t)/a$ is the Hubble parameter. Horizon entry therefore occurs when $k^{-1}= (H^{-1})/a$, or equivalently $k=a H$.
The size of a density perturbation is parameterised using the density contrast, $\delta(M_{\rm H})$, defined as
\begin{equation}
\delta(M_{\rm H}) \equiv \frac{\delta \epsilon}{\epsilon_{\rm tot}} \equiv \frac{ \epsilon(M_{\rm H})-\epsilon_{\rm tot}}{\epsilon_{\rm tot}} \,,  
\end{equation}
where $\epsilon(M_{\rm H})$ is the density of the region and $\epsilon_{\rm tot}$ is the mean density of the universe.  We have
parameterised the scale of the region using the horizon mass, the mass within the horizon radius, $H^{-1}$:
\begin{equation}
 M_{\rm H} = \frac{4}{3} \pi \epsilon_{\rm tot} (H^{-1})^{3} \,.
\end{equation}
During radiation domination $\epsilon_{\rm tot} = \epsilon_{\rm r} \propto a^{-4}$, $a \propto t^{1/2}$ and $H \propto t^{-1}$, so that $M_{\rm H} \propto a^{2} \propto t$. Using this scaling we can express the horizon mass during radiation domination in terms of the horizon mass at matter-radiation equality (when the matter and radiation densities are equal), $M_{\rm eq}$,:
\begin{equation}
\label{mh}
M_{\rm H} \approx \left( \frac{t}{t_{\rm eq}} \right) M_{\rm eq} \,,
\end{equation}
where $M_{\rm eq} \approx 1 \times 10^{17} \, M_{\odot}$ and $t_{\rm eq} \approx 2 \times 10^{12} \, {\rm s}$.

In early work, Carr showed that a perturbation would collapse to form a PBH if its amplitude at horizon entry, $\delta(M_{\rm H})$, exceeds a critical, threshold, value $\delta_{\rm c}$~\cite{Carr:1975qj}. He found that $\delta_{\rm c} = c_{\rm s}^2$, where $c_{\rm s}$ is the sound speed (which determines how quickly pressure waves can traverse the perturbation). During radiation domination $c_{\rm s}^2 = w =1/3$, where $w$ is the equation of state parameter which relates density and pressure, $p$: $w = p/\epsilon$. More recently, numerical simulations have found $\delta_{\rm c}$=0.45, for a review see Ref.~\cite{Escriva:2021aeh}. The mass of the resulting PBH, $M$, is roughly equal to the horizon mass, $M_{\rm H}$, and can be written, using Eq.~(\ref{mh}), as
\begin{equation}
M \sim 10^{15} \, {\rm g} \left( \frac{t}{10^{-23}} \right) \sim  M_{\odot} \left( \frac{t}{10^{-6} \, {\rm s}} \right) \,.
\end{equation}
Here we have normalised firstly to the (initial) mass of PBHs that would be completing their evaporation today, $M \sim 10^{15} \, {\rm g}$ (see Sec.~\ref{sec-evap}) and secondly to the time, $t \approx 10^{-6} \, {\rm s}$, of the QCD phase transition (when the universe transitions from a quark-gluon plasma to a hadronic phase) when the horizon mass is of order a Solar mass, $M_{\odot}$. The reduction in pressure at phase transitions leads to a reduction in the threshold for PBH formation, $\delta_{\rm c}$, so that PBHs form more easily on scales that correspond to phase transitions. 

In the late 1990s, it was pointed out that, due to critical phenomena, the PBH mass depends on the size of the fluctuation from which it forms~\cite{Niemeyer:1997mt,Jedamzik:1999am}: 
\begin{equation}
M = k M_{\rm H} \left(\delta(M_{\rm H}) - \delta_{\rm c} \right)^{\gamma} \,,
\end{equation}
where $k$ is a constant of order unity and $\gamma \approx 0.36$. PBHs formed at the same time have a range of masses, however most have mass within roughly an order of magnitude of $M_{\rm H}$~\cite{Jedamzik:1999am}. More recently it has been realised that the criterion for PBH formation is best expressed in terms of the peak value of the compaction function, which is the average mass excess in a given volume. For further details see e.g. Ref.~\cite{Escriva:2022duf}.

\subsection{Abundance}
\label{sec-abundance}

In this Section, we will first outline an approximate, illustrative calculation of the abundance of PBHs. We will then see how to translate the initial abundance into the present-day PBH density. Finally, we will briefly overview refinements to this simple calculation.

\begin{figure}[t]
	\centering
	\includegraphics[width=.8\textwidth]{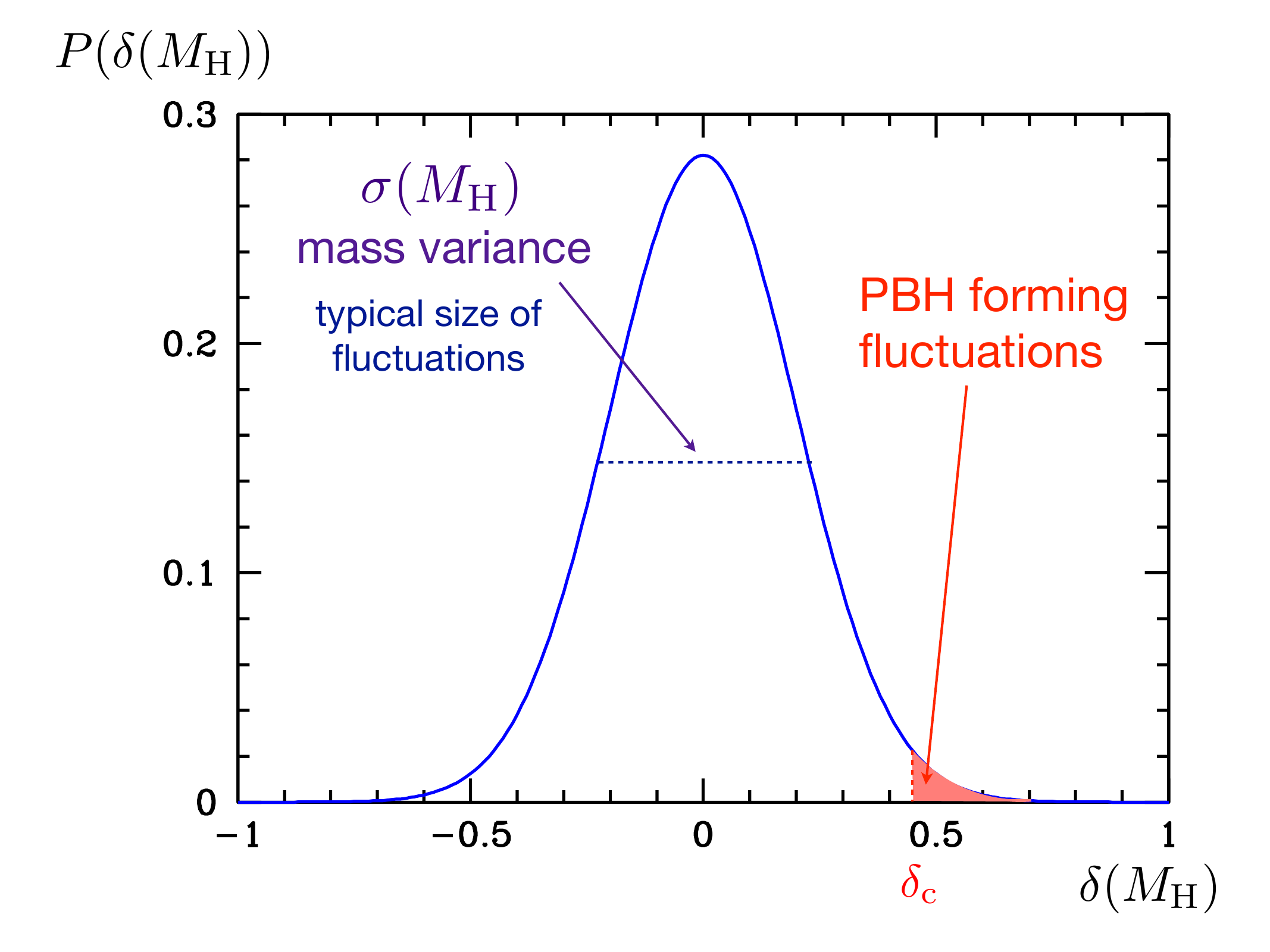}
	\caption{A schematic illustration of the probability distribution, $P(\delta(M_{\rm H}))$, of the size of the density perturbations, $\delta(M_{\rm H})$, on a particular scale denoted by the horizon mass $M_{\rm H}$. PBHs form from rare large fluctuations with size greater than the critical value, $\delta_{\rm c}$, which is of order $0.45$. The width of the distribution, or equivalently the typical size of the fluctuations, is given by the mass variance, $\sigma(M_{\rm H})$. This diagram is not to scale; as discussed in the text, the fraction of perturbations which collapse to form PBHs has to be very small and hence $\sigma(M_{\rm H})$ has to be much smaller than $\delta_{\rm c}$.}
	\label{fig:prob}
\end{figure}

As we saw in Sec.~\ref{sec-criteria}, a fluctuation will collapse to form a PBH if the density contrast at horizon entry (on a particular scale, parameterised by the horizon mass, $M_{\rm H}$), $\delta(M_{\rm H})$, exceeds a critical value, $\delta_{\rm c}$, which is of order $0.45$. The initial abundance of PBHs is usually quantified in terms of the initial mass fraction, $\beta(M)$,
\begin{equation}
\beta(M) \equiv \left( \frac{\epsilon_{\rm PBH}}{\epsilon_{\rm tot}} \right)_{\rm i} \,,
\end{equation}
where $\epsilon_{\rm PBH}$ is the PBH energy density and `i' denotes the formation time.
Assuming for simplicity that all PBH form at the same time and have mass equal to the horizon mass, $\beta(M)$ is roughly equal to the fraction of the Universe, at the formation time, which has a density contrast larger than $\delta_{\rm c}$ (see Fig.~\ref{fig:prob} for a schematic illustration): 
\begin{equation}
\beta(M) \approx \int_{\delta_{\rm c}}^{\infty} P(\delta(M_{\rm H})) \, {\rm d} \delta(M_{\rm H}) \,, 
\end{equation}
where $P(\delta(M_{\rm H})$ is the probability distribution of the density contrast, $\delta(M_{\rm H})$. Assuming a Gaussian probability distribution, with width $\sigma(M_{\rm H})$ (known as the mass variance)
\begin{equation}
P(\delta(M_{\rm H})) \, {\rm d}  \delta(M_{\rm H})  =  \frac{1}{\sqrt{2 \pi} \sigma(M_{\rm H})} \exp{  \left( - \frac{\delta^2 }{2 \sigma^2(M_{\rm H}) } \right)  } {\rm d} \delta(M_{\rm H})  \,,
\end{equation} 
then the mass fraction can be written in terms of the complementary error function defined as
\begin{equation}
{\rm erfc} (x) = \frac{2}{\sqrt{\pi}} \int_{x}^{\infty} \exp{(-z^2)} \, {\rm d} z \,,
\end{equation} 
as 
\begin{equation}
\beta(M) \approx \frac{1}{2} {\rm erfc} \left( \frac{ \delta_{\rm c}}{\sqrt{2} \sigma (M_{\rm H})} \right) \,.
\end{equation}
We will shortly see that the initial PBH abundance has to be very small, $\beta \ll 1$. This requires $\sigma(M_{\rm H}) \ll \delta_{\rm c}$, i.e.~only a tiny fraction of the Universe can form PBHs. In this limit, using the large argument limit of the complementary error function, ${\rm erfc} (x) \approx \exp{(-x^2)}/ ( x \sqrt{\pi})$,
\begin{equation}
\label{beta}
\beta(M) \approx \frac{\sigma(M_{\rm H})}{\sqrt{2 \pi} \delta_{\rm c} } \exp{ \left( - \frac{\delta_{\rm c}^2}{2 \sigma^2(M_{\rm H})} \right)} \,.
\end{equation}
The size of the primordial perturbations is usually parameterised in terms of the power spectrum of the curvature perturbation, ${\cal R}$,  ${\cal P}_{\cal R} = [k^3/ (2 \pi^2)] |\tilde{R}(k)|^2 $ where $\tilde{R}(k)$ is the Fourier transform of the curvature perturbation. The curvature perturbation determines the perturbation in the total energy density, for a formal definition see e.g.~Refs.~\cite{Byrnes:2021jka,Baumann:2022mni}.
The mass variance is calculated by integrating the dimensionless power spectrum of the density contrast multiplied by the square of the Fourier transform of the window function used to smooth the density contrast, see e.g.~Ref.~\cite{Green:2020jor}. However, to within an order of magnitude,  
\begin{equation}
\label{sigma}
\sigma^2(M_{\rm H})  \sim {\cal P}_{\cal R}(k) \,.
\end{equation}
During radiation domination, the comoving wavenumber corresponding to the horizon, $k = a H$, scales as $k \propto t \propto a$. Using the scaling of the horizon mass from Sec.~\ref{sec-criteria}, $M_{\rm H} \propto k^{-2}$ and using the horizon mass and comoving wavenumber corresponding to the Hubble radius, $k_{\rm eq} =10^{-2} \, {\rm Mpc}^{-1}$, at matter-radiation equality
\begin{equation}
M_{\rm H} \sim 10^{13} M_{\odot} \left(   \frac{1 \, {\rm Mpc}^{-1}}{k}   \right)^2  \,.
\end{equation}

\begin{figure}[t]
	\centering
	\includegraphics[width=.8\textwidth]{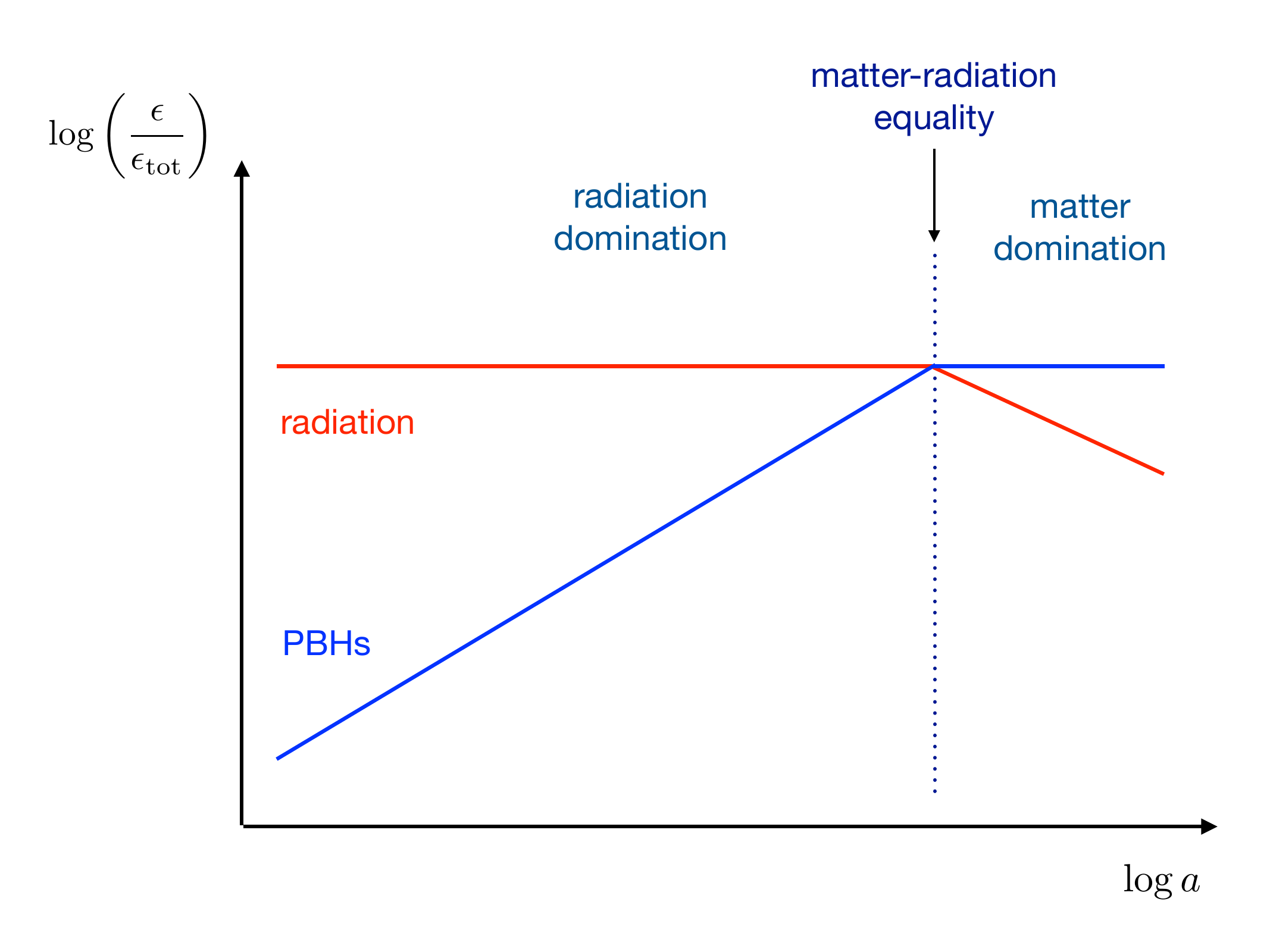}
	\caption{A schematic illustration of the fraction of the total energy density in the form of radiation (in red) and PBHs (blue) as a function of the scale factor $a$. The dashed vertical line indicates the epoch of matter-radiation equality, where the Universe transitions from being dominated by radiation to dominated by matter. PBHs are matter so their energy density varies with the scale factor, $a$, as $\epsilon_{\rm PBH} \propto a^{-3}$, while the density of radiation varies as $\epsilon_{\rm r} \propto a^{-4}$. Consequently the fraction of the energy density in the form of PBHs, $\epsilon_{\rm PBH}/ \epsilon_{\rm tot}$ is constant during matter domination and grows proportional to $a$ during radiation domination. }
	\label{fig:density}
\end{figure}

Since PBHs are massive, and hence non-relativistic, their density is inversely proportional to volume $\epsilon_{\rm PBH} \propto a^{-3}$. During radiation domination, the total energy density varies as $\epsilon_{\rm tot} = \epsilon_{\rm r} \propto a^{-4}$ and hence the fraction of the density in PBHs grows proportional to $a$, while during matter domination PBHs make up a constant fraction of the total density (see Fig.~\ref{fig:density}). We will now relate the initial PBH abundance, $\beta(M)$, to $f_{\rm PBH}$, the fraction of the DM in the form of PBHs today (denoted by subscript `0'),
\begin{equation}
f_{\rm PBH} \equiv \left( \frac{ \epsilon_{\rm PBH}}{\epsilon_{\rm DM}} \right)_{0} \,.
\end{equation}
Using the variation of the matter and radiation densities with the scale factor we can relate the PBH and radiation densities at the time of PBH formation (denoted by `i' for initial)  with those at matter-radiation equality (denoted by `eq'),
\begin{equation}
\epsilon_{{\rm PBH, i}} = \epsilon_{{\rm PBH, eq} } \left( \frac{a_{\rm eq}}{a_{\rm i}} \right)^3 \,, \hspace{1.0cm} \epsilon_{{\rm r, i}} = \epsilon_{{\rm r, eq} } \left( \frac{a_{\rm eq}}{a_{\rm i}} \right)^4  \,, \nonumber 
\end{equation}
so that 
\begin{equation}
\beta(M)  \equiv  \left( \frac{\epsilon_{\rm PBH}}{\epsilon_{\rm tot}} \right)_{\rm i} =  \left( \frac{\epsilon_{\rm PBH}}{\epsilon_{\rm r}} \right)_{\rm eq}  \left( \frac{a_{\rm i}}{a_{\rm eq}} \right) \,. \nonumber 
\end{equation}
Using the fact that the matter and radiation densities are equal at matter-radiation equality, assuming that the Universe is matter-dominated from matter-radiation equality to the present day and neglecting the contribution of baryons to the matter density: 
\begin{equation}
f_{\rm PBH} \equiv \left( \frac{ \epsilon_{\rm PBH}}{\epsilon_{\rm DM}} \right)_{0}  = \left( \frac{ \epsilon_{\rm PBH}}{\epsilon_{\rm DM}} \right)_{{\rm eq}}  = \left( \frac{\epsilon_{\rm PBH}}{\epsilon_{\rm r}} \right)_{\rm eq} =  \left( \frac{a_{\rm eq}}{a_{\rm i}} \right)  \left( \frac{\epsilon_{\rm PBH}}{\epsilon_{\rm tot}} \right)_{\rm i} = \left( \frac{a_{\rm eq}}{a_{\rm i}} \right) \, \beta(M) \,.
\end{equation}
During radiation domination, $a \propto t^{1/2} \propto M^{1/2}$ (see Sec.~\ref{sec-criteria}) and using the horizon mass at matter-radiation equality
\begin{equation}
   \left( \frac{a_{\rm i}}{a_{\rm eq}} \right) =   \left( \frac{M_{{\rm }}}{M_{{\rm eq}}} \right)^{1/2}  = 4 \times 10^{-10} \left(  \frac{M_{{\rm }}}{M_{{\odot}}} \right)^{1/2} \,, \nonumber 
 \end{equation}   
and hence
\begin{equation}
\label{fbeta}
f_{\rm PBH} \sim 10^{9} \beta(M_{\rm H}) \left(  \frac{M_{{\rm }}}{M_{{\odot}}} \right)^{-1/2} \,.
\end{equation} 
For a more accurate expression,
see e.g.~Ref.~\cite{Tomberg:2024chk}.

PBHs with initial mass $M \gtrsim 10^{15} \, {\rm g}$ are thought to be stable (see Sec.~\ref{sec-evap}). Their present-day density can not exceed the measured DM density, $f_{\rm PBH} \leq 1$ and hence the initial mass fraction $\beta(M_{\rm H})$ must (as previously mentioned) be small. If, for instance, Solar mass PBHs make up all of the DM then we require $\beta(1 M_{\odot}) \sim 10^{-9}$. Lighter PBHs form earlier, and therefore their density grows relative to the background density for longer, and hence their initial density must be even smaller.

On cosmological scales the amplitude of the dimensionless power spectrum has been measured to be ${\cal P}_{\cal R} \sim 10^{-10}$~\cite{Planck:2018jri}. If the perturbations are scale-invariant, i.e.~have the same amplitude on all scales, then $\sigma \sim {\cal P}_{\cal R}^{1/2} \sim 10^{-5}$. Using Eq.~(\ref{beta})  in this case
\begin{equation}
\beta(M_{\rm H}) \sim {\rm erfc} \left(10^{5} \right) \sim \exp{ \left( -10^{10} \right) } \lll 1 \,,
\end{equation}
i.e.~the number of PBHs formed is completely negligible. How large does the mass variance have to be for PBHs to make up a non-negligible fraction of the DM? Inverting Eq.~(\ref{beta})  
\begin{equation}
\sigma(M_{\rm H}) \sim \frac{\delta_{\rm c}}{\left[ -\ln{\beta(M_{\rm H})} \right]^{1/2}} \,.
\end{equation}
We saw above that for Solar mass PBHs to make up all of the DM, $\beta(1 M_{\odot}) \sim 10^{-9}$, and hence $\sigma(1 M_{\odot}) \sim 10^{-1}$ i.e.~the amplitude of the perturbations has to be several orders of magnitude larger than the measured value on cosmological scales. The value of $\beta(M_{\rm H})$ corresponding to a fixed value of $f_{\rm PBH}$ is strongly dependent on the PBH mass, see Eq.~(\ref{fbeta}). However, because of the exponential dependence of the PBH abundance on the amplitude of the perturbations, the corresponding value of $\sigma(M_{\rm H})$ only varies by a factor of a few. This exponential dependence also means that the value of $\sigma(M_{\rm H})$ has to be fine-tuned to obtain a particular value of $f_{\rm PBH}$.

Before looking at the generation of large density perturbations by inflation in Sec.~\ref{sec-inflation}, we conclude this section by mentioning some important caveats to the simple illustrative calculations above which must be taken into account when doing accurate calculations. We have assumed that all PBHs form at the same time and have the same mass. 
As mentioned in Sec.~\ref{sec-criteria}, critical phenomena mean that PBHs formed at a fixed time have a range of masses. It is also possible that fluctuations are sufficiently large to form PBHs on a range of scales, and hence at a range of times. To accurately calculate the present-day abundance, and mass function, of PBHs both the spread in formation times and the spread in masses of PBHs formed at a given time must be taken into account (see e.g.~Ref.~\cite{Germani:2023ojx}). When estimating the initial abundance of PBHs, Eq.~(\ref{beta}), we assumed that the probability distribution of the density fluctuations is Gaussian. Since PBHs form from rare large fluctuations, in the tail of the probability distribution, the abundance of PBHs formed depends sensitively on the shape of the probability distribution, and the probability distribution of large density perturbations is expected to be non-gaussian (for further details see e.g. Ref.~\cite{Byrnes:2021jka}).  Finally, we noted in Sec.~\ref{sec-criteria} that the reduction in pressure at phase transitions leads to a reduction in the threshold for collapse, $\delta_{\rm c}$, on the corresponding scales. Consequently, for a fixed amplitude of the density contrast, PBHs with mass corresponding to the horizon mass at the time of the phase transition will form more abundantly.

\subsection{Generation of large density perturbations by inflation}
\label{sec-inflation}

Inflation is a period of accelerated expansion postulated to have occurred in the very early Universe, in the first tiny fraction of a second after the Big Bang. It is typically driven by a scalar field (a field with only a single value at each point in space), $\phi$, evolving slowly along a potential, $V(\phi)$, which is close to flat (referred to as `slow roll'). Inflation was proposed in the 1980s to solve various problems with the standard Hot Big Bang~\cite{Starobinsky:1980te,Guth:1980zm,Sato:1980yn}. Quantum fluctuations in the scalar field generate density perturbations~\cite{Guth:1982ec,Hawking:1982cz,Linde:1982uu}, from which structures (galaxies, galaxy clusters etc.) can later form. 

The density perturbations also generate anisotropies in the temperature of the Cosmic Microwave Background (CMB). The temperature fluctuations in the CMB, as measured by Planck and other experiments, are consistent with the predictions of the simplest single field slow-roll inflation models~\cite{Planck:2018jri}. In particular, on the length scales probed by the   
CMB (comoving wavenumbers, $k$, in the range $0.001$ to $0.1 \, {\rm Mpc}^{-1}$) the distribution of the sizes of the perturbations is close to Gaussian, and the typical size of the perturbations is of order $10^{-5}$, with a small scale dependence. More precisely, on cosmological scales the power spectrum of the curvature perturbation (which is roughly equal to the square of the typical size of the fluctuations, $\sigma$, see Sec.~\ref{sec-abundance}) is well fit by
\begin{equation}
\label{pkcmb}
 {\cal P}_{\cal R}(k) = 2 \times 10^{-9}  \left( \frac{k}{0.05 \, {\rm Mpc}^{-1}} \right)^{n_{\rm s} -1}  \,,
\end{equation}
with spectral index $n_{\rm s} = 0.963 \pm 0.006 $~\cite{Planck:2018jri}.
As we saw in Sec.~\ref{sec-criteria}, the density perturbations on shorter length scales can form PBHs provided their amplitude is sufficiently large. 
However if the perturbations have roughly the same amplitude on small scales as on cosmological scales then the abundance of PBHs formed is completely negligible. To form an interesting abundance of PBHs the perturbations on small scales (large $k$) must be several orders of magnitude larger.

Here we give a brief overview of the aspects of inflation that are important for generating large, PBH forming, density perturbations, for a more detailed review see e.g.~Ref.~\cite{Byrnes:2021jka}.
A scalar field has density, $\epsilon$, and pressure, $p$ given by
\begin{equation}
\epsilon = \frac{1}{2} \dot{\phi}^2 + V(\phi) \,, \hspace{1.0cm} p = \frac{1}{2} \dot{\phi}^2 - V(\phi) \,,
\end{equation}
where $\dot{}$ denotes a derivative with respect to time.
The Friedmann equation for the evolution of a universe dominated by a scalar field is
\begin{equation}
\label{friedmann}
H^2 \equiv \left( \frac{\dot{a}}{a} \right)^2 = \frac{1}{3 M_{\rm Pl}^2} \left[ \frac{1}{2} \dot{\phi}^2 + V(\phi) \right] \,,
\end{equation}
where $H$ is the Hubble parameter (i.e.~the expansion rate of the universe) and $M_{\rm Pl}= 1/(8 \pi G)$ is the reduced Planck mass. The evolution of the scalar field is governed by the Klein-Gordon equation
\begin{equation}
\label{KG} 
\ddot{\phi} + 3 H \dot{\phi} +  \frac{{\rm d} V}{{\rm d} \phi}  = 0 \,.
\end{equation}

The dynamics of inflation can be described effectively using the (first two) Hubble slow-roll parameters, $\epsilon_{H}$ and $\eta_{H}$:
\begin{equation}
\epsilon_{H} \equiv - \frac{\dot{H}}{H^2} \,, \hspace{1.0cm}  \eta_{H} \equiv \frac{ \dot{\epsilon}_{H} }{H \epsilon_{H}} \,.
\end{equation}
Inflation occurs, $\ddot{a} > 0$, when $\epsilon_{H} <1$. 
In the slow roll approximation $\epsilon_{H} \ll 1$ and $|\eta_{H}| \ll 1$ and the Friedman and Klein-Gordon equations, Eqs.~(\ref{friedmann}) and (\ref{KG}), become 
\begin{equation}
H^2 \approx \frac{V}{3 M_{\rm Pl}^2} \,, \hspace{1.0cm}  3 H \dot{\phi} \approx -  \frac{{\rm d} V}{{\rm d} \phi} \,,
\end{equation}
i.e. the dynamics of the scalar field are like a ball rolling down a hill, with the expansion of the universe providing friction. In the limit that $\epsilon_{H} = 0 $, $H$ is constant and the universe expands (exactly) exponentially.

\begin{figure}[t]
	\centering
	\includegraphics[width=.8\textwidth]{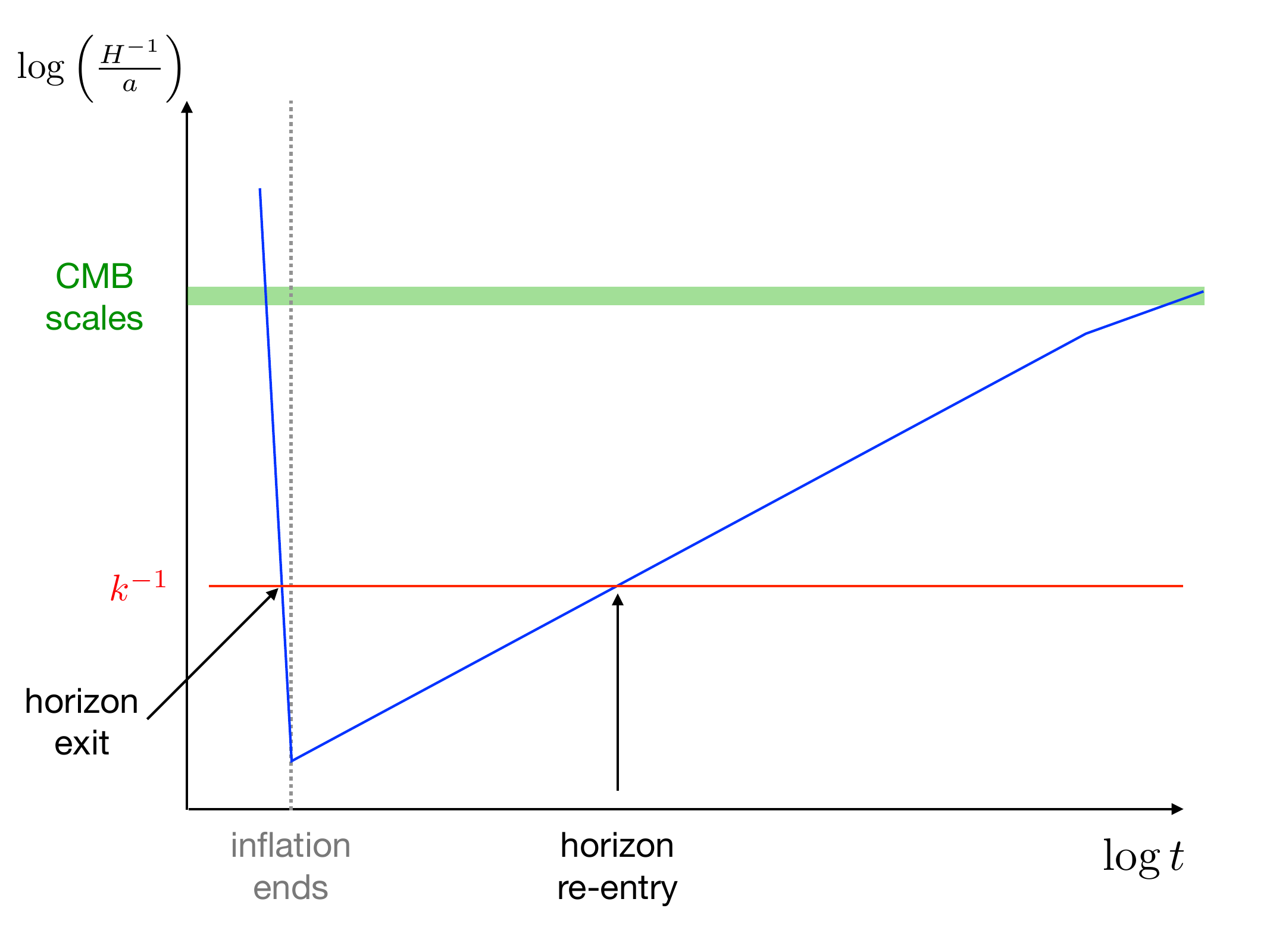}
	\caption{A schematic illustration of the variation of the comoving Hubble radius $H^{-1}/a$ with time, $t$. During inflation it decreases roughly exponentially (since $H$ is roughly constant and $a$ grows roughly exponentially), while during the subsequent radiation and matter-dominated eras it grows as $t^{1/2}$ and $t^{1/3}$ respectively. A given scale, with comoving wavenumber $k$, (shown as the horizontal red line) exits the horizon when $k=aH$ during inflation. It is then `super-horizon' until it re-enters the horizon when $k=aH$ again during radiation or matter domination. Small physical scales have large $k$ and hence, compared to cosmological scales, they exit the horizon later and re-enter earlier. If a density fluctuation that re-enters the horizon during radiation domination is sufficiently large it will collapse to form a PBH soon after horizon re-entry (see Sec.~\ref{sec-criteria}). The shaded region corresponds to cosmological scales.
	\label{fig:horizon}}
\end{figure}

It can be shown that the power spectrum of the curvature perturbation is given by
\begin{equation}
{\cal P}_{\cal R} = \frac{1}{8 \pi^2 M_{\rm Pl}^2} \frac{H_{\star}^2}{\epsilon_{H, \star}} \,,
\end{equation}
where $\star$ denotes that the variables should be evaluated when the scale exits the horizon ($k= a H$) during inflation. During matter and radiation domination, the comoving Hubble radius $(H^{-1})/a$ grows with time, however during inflation it decreases roughly exponentially, see Fig.~\ref{fig:horizon}.

In the slow-roll limit $H^2 \propto V$ and $\epsilon_{H} \propto (V^{\prime})^2/ V^2$ and hence ${\cal P}_{\cal R} \propto V^3/(V^{\prime})^2$. This indicates that to obtain large ${\cal P}_{\cal R}$ we require small $V^{\prime}$ i.e.~an inflexion point or a small local minimum in the potential. In this case, known as `ultra slow-roll' (USR) inflation, the $V^{\prime}$ term in the Klein Gordon equation, Eq.~(\ref{KG}), is negligible rather than the $\ddot{\phi}$, term and the motion of the scalar field is governed by
\begin{equation}
 \ddot{\phi} + 3 H \dot{\phi} \approx  0 \,.
\end{equation}
This equation has solution $\dot{\phi} \propto a^{-3}$, $\epsilon_{H} \propto a^{-6} \ll 1$ and $\eta_{H} \approx -6$. Since $\epsilon_{H}$ decreases very rapidly (while $H$ is roughly constant), ${\cal P}_{\cal R}$ grows rapidly. Note that (despite its name) ultra slow-roll is not a limiting case of slow-roll, since $|\eta_{H}| \nll 1$ the slow-roll approximation is in fact not valid. This has various consequences, see e.g.~Ref.~\cite{Byrnes:2021jka}. In particular, the curvature perturbations continue growing after horizon exit (which is not the case for slow-roll inflation) and a numerical calculation is required to accurately evaluate the power spectrum.

\begin{figure}[t]
	\centering
	\includegraphics[width=.8\textwidth]{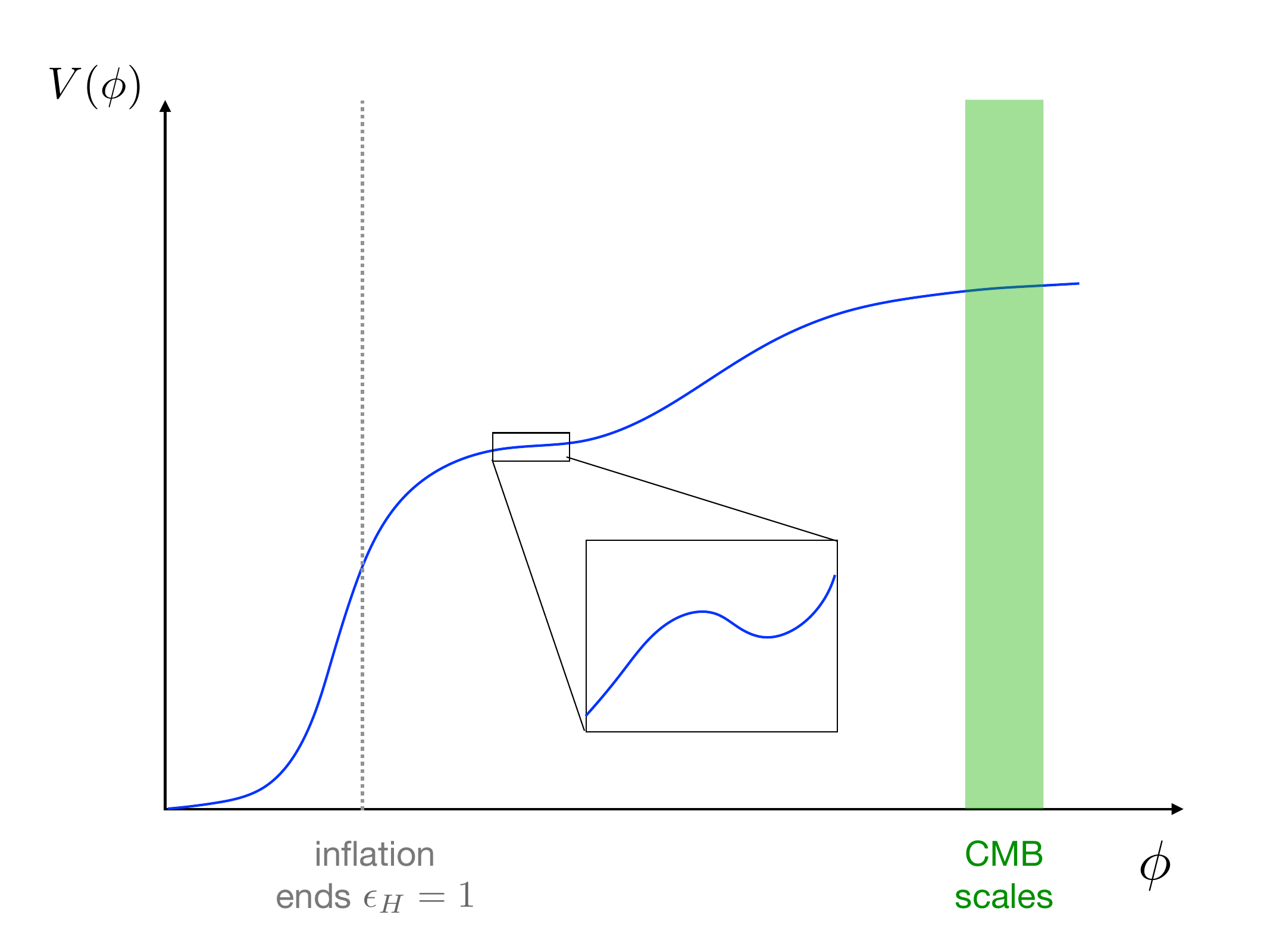}
	\caption{A schematic illustration of an inflaton potential, $V(\phi)$, that can produce large, PBH forming perturbations. The shaded region corresponds to cosmological scales on which the amplitude and scale dependence of the primordial power spectrum are accurately measured. The field slows down when it reaches the small feature, and inflation ends when the potential steepens so that $\epsilon_{H}=1$. 
	\label{fig:potential}}
\end{figure}

\begin{figure}[t]
	\centering
	\includegraphics[width=.8\textwidth]{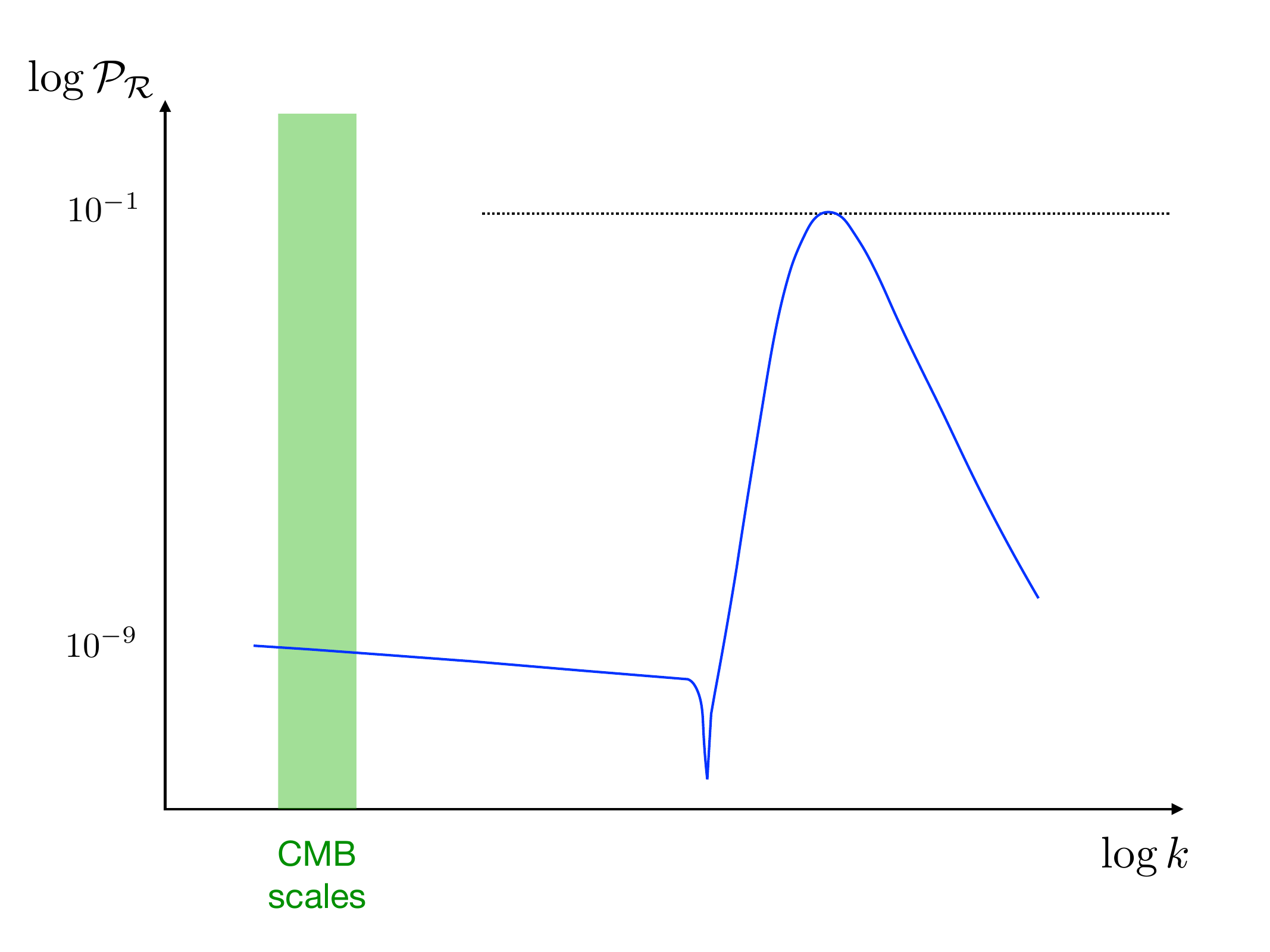}
	\caption{A schematic illustration of the power spectrum of the curvature perturbation, ${\cal P}_{\cal R}(k)$, produced by an inflation model with a small feature in the potential, as in Fig.~\ref{fig:potential}, which leads to a period of ultra-slow roll inflation and a large enhancement in the amplitude of the perturbations. The shaded region corresponds to cosmological scales on which the amplitude and scale dependence of the primordial power spectrum are accurately measured. The dashed horizonal line shows the approximate amplitude required to generate an interesting abundance of PBHs, ${\cal P}_{\cal R}(k) \sim 0.1$.
	\label{fig:powerspectrum}}
\end{figure}

An inflation model which successfully produces PBHs has to satisfy three requirements:
\begin{enumerate}
\item{The amplitude and scale-dependence (i.e.~the spectral index) of the power spectrum on cosmological scales are consistent with CMB observations, Eq.~(\ref{pkcmb}).}
\item{The amplitude of the power spectrum is several orders of magnitude larger, ${\cal P}_{\cal R} \sim 0.1$, on some smaller scale.}
\item{Inflation subsequently ends, $\epsilon_{H} > 1$. }
\end{enumerate}
It is possible to satisfy all three requirements in single-field models, with a small feature in the potential, however fine-tuning is required. Figure \ref{fig:potential} shows an example of a potential with a small feature which slows the inflaton field down, leading to a large enhancement in the power spectrum as shown in Fig.~\ref{fig:powerspectrum}. The three requirements can be more easily satisfied in multi-field inflation models where, for instance, a different field is responsible for generating the perturbations than for ending or driving inflation. A particular example is hybrid inflation, where inflation ends due to a second field undergoing a phase transition at which point large quantum fluctuations can be generated.
It should be noted that while there are various inflation models that produce perturbations large enough to generate an interesting abundance of PBHs, this is not a generic feature of inflation (i.e.~the vast majority of inflation models do not generate large PBH forming perturbations). For reviews of PBH-producing inflation models see Refs.~\cite{Escriva:2022duf,Ozsoy:2023ryl}


\subsection{Other mechanisms}
\label{sec-othermech}

PBHs can also form from the collapse of density perturbations during matter domination, however in this case the perturbation has to be sufficiently spherical and homogeneous.  
There are various other mechanisms via which PBHs may form in the early Universe, including the collapse of cosmic string loops, collisions of bubbles formed during first-order phase transitions, the collapse of domain walls produced at a second-order phase transition, the fragmentation of a scalar field condensate and via long-range fifth forces.
For more detailed overviews of these PBH formation mechanisms, and references, see Refs.~\cite{Green:2020jor,Carr:2021bzv,Escriva:2022duf}.

\section{Observational probes of Primordial Black Holes} \label{sec-obs}

\begin{figure}[t]
	\centering
	\includegraphics[width=.8\textwidth]{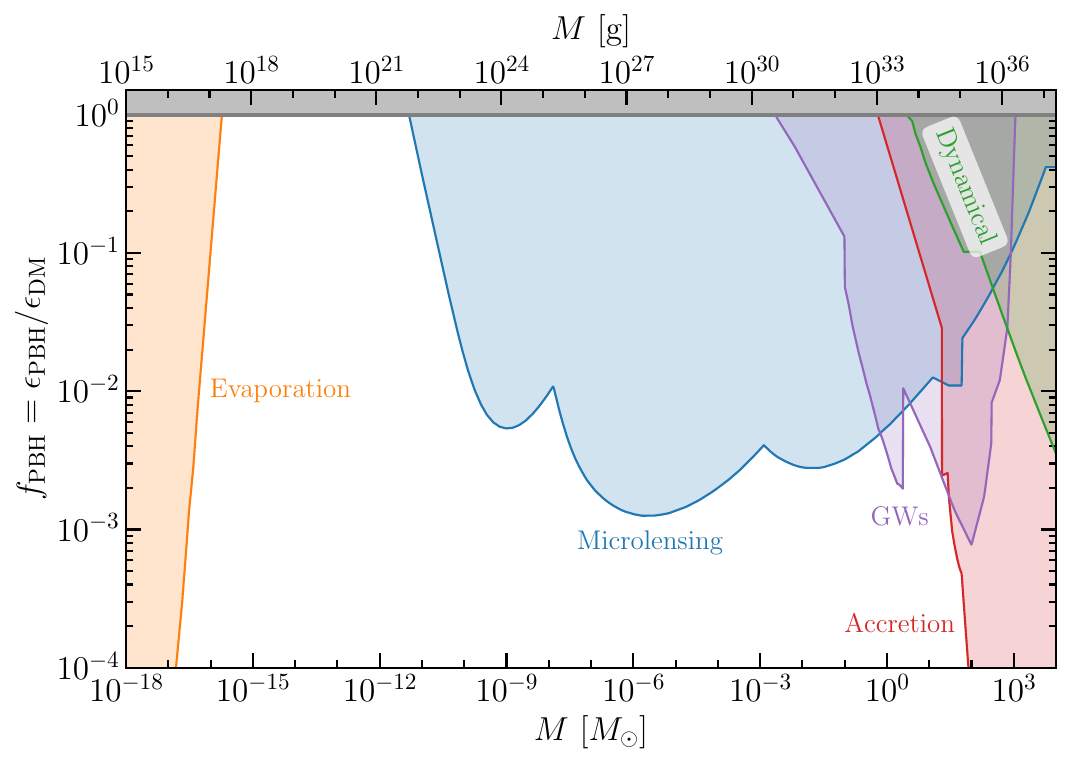}
	\caption{The constraints (as of November 2024) on the fraction of DM in the form of PBHs, $f_{\rm PBH}$, as a function of mass, $M$, assuming all PBHs have the same mass. The bounds shown are (left to right) from evaporation (orange), microlensing (blue), gravitational waves (purple), accretion (red) and dynamical effects (green). For each type of bound the tightest constraint at each mass is shown and the shaded regions are excluded under standard assumptions. Created using Kavanagh’s PBHbounds code https://github.com/bradkav/PBHbounds. 
	\label{fig:constraints}}
\end{figure}

In this Section, we overview observational probes of the abundance of PBHs. There are a large number of different constraints. In this Encyclopedia entry, we focus on those that are currently tightest. We start with a general introduction in Sec.~\ref{sec-obsintro}. Then we discuss evaporation constraints on light ($M \lesssim 10^{17} \, {\rm g}$) PBHs (Sec.~\ref{sec-evap}), stellar microlensing constraints on planetary to multi-Solar mass PBHs (Sec.~\ref{sec-lensing}) and other direct constraints (Sec.~\ref{sec-other}). In Sec.~\ref{sec-indirect} we cover constraints on the amplitude of density perturbations, which place indirect limits on the abundance of PBHs formed via the collapse of large density perturbations. Finally, we conclude in Sec.~\ref{sec-asteroid} with a discussion of ideas for probing currently unconstrained asteroid mass PBHs. Figure \ref{fig:constraints} shows the tightest constraint on $f_{\rm PBH}$ as a function of $M$, assuming all PBHs have the same mass.

\subsection{Introduction}
\label{sec-obsintro}

Comparing theoretical models with observations inevitably involves assumptions and modelling. The validity of, and uncertainties in, constraints (or best-fit parameter values in the case of a positive signal) depends on the accuracy of the assumptions and modelling. We will now briefly discuss the assumptions that are typically made when calculating observational constraints on the abundance of PBHs.

Constraints on the abundance of PBHs are usually calculated assuming a delta-function mass function, i.e.~that PBHs all have the same mass. As discussed in Sec.~\ref{sec-criteria}, due to critical phenomena in gravitational collapse, the mass function is expected to have a finite width. The constraint for a given extended mass function can (in most cases) be calculated from the delta-function constraint using the method presented in Ref.~\cite{Carr:2017jsz}. For a specific constraint, the tightest limit on $f_{\rm PBH}$ is weaker for an extended mass function than for a delta function, however, $f_{\rm PBH}$ is less than one for a wider range of masses. Consequently, the range of mass for which $f_{\rm PBH}=1$ is allowed becomes smaller, i.e.~when all constraints are considered, extended mass functions are more tightly constrained 

Many constraints also depend on the spatial distribution of PBHs. PBHs formed from the collapse of large density perturbations are more clustered on sub-galactic scales than particle dark matter, due to the Poisson fluctuations in their initial distribution~\cite{Afshordi:2003zb}. The extent to which clustering affects constraints depends on how compact the clusters are~\cite{DeLuca:2022uvz}. PBHs formed from Gaussian density perturbations form relatively diffuse clusters~\cite{Inman:2019wvr} but PBHs formed from non-Gaussian perturbations (or via other mechanisms) could form more compact clusters~\cite{Young:2019gfc}.

Finally, there are some observations or phenomena which could potentially be due to PBHs (for a review see Ref.~\cite{Carr:2023tpt}). Assuming PBH are responsible for an observation, the corresponding PBH abundance and mass function can then be calculated. However, there may be other potential explanations of these phenomena, and a PBH interpretation may be inconsistent with exclusion limits from other observations. 

\subsection{Hawking evaporation}
\label{sec-evap}

Hawking \cite{Hawking:1974rv, Hawking:1975vcx} showed that a non-rotating, uncharged BH with mass $M$ radiates with a temperature, $T$, given by
\begin{equation}
    k_{\rm B} T = \frac{\hbar}{8\pi G M} = 1.06 \left(\frac{M}{10^{16} \, {\rm g}}\right)^{-1} \, {\rm MeV} \,,
\end{equation}
where $G$ is the gravitational constant, $\hbar$ is the reduced Planck constant and $k_{\rm B}$ is the Boltzmann constant. 
BHs can directly emit particle species with rest mass energy less than $k_{\rm B} T$ and 
 the BH mass decreases at a rate \cite{Page:1976df,MacGibbon:1991vc}
\begin{equation}
\label{dmdt}
    \frac{{\rm d} M}{{\rm d} t} \approx  5 \times 10^{25} f(M) \left( \frac{M}{1 \, {\rm g}} \right)^{-2} \, {\rm g} \, {\rm s}^{-1} \,,
\end{equation}
where $f(M)$ depends on the number of species that can be emitted and is normalized to one for BHs with mass $M \gg 10^{17} \, {\rm g}$ that can only emit massless particles.
Taking into account the increase in $f(M)$ as $M$ decreases, PBHs with initial mass $M \sim 10^{15} \, {\rm g}$ are completing their evaporation today, and PBHs more massive than this can (in principle) be the DM.

PBHs with initial mass $10^{15} \, {\rm g} \lesssim M \lesssim 10^{17} \, {\rm g}$ are evaporating at a significant rate, and their abundance can be constrained using limits on the flux of the particles they produce, for instance, MeV gamma-rays. Since the PBH evaporation rate decreases rapidly with increasing PBH mass (Eq.~(\ref{dmdt})) the constraints weaken rapidly with increasing $M$: $f_{\rm PBH} \lesssim 10^{-4}$ for $M \approx 10^{16} \, {\rm g}$ weakening to $f_{\rm PBH} \lesssim 1$ for $M \approx 10^{17} \, {\rm g}$.  For a recent review of PBH evaporation and the resulting constraints on the abundance of PBHs, see Ref.~\cite{Auffinger:2022khh}.

\subsection{Stellar microlensing}
\label{sec-lensing}

Stellar microlensing is the temporary brightening, due to gravitational lensing, of a star which occurs when a compact object (CO) traverses the line-of-sight between the observer and the star~\cite{Paczynski:1985jf,Griest:1990vu}. It is referred to as microlensing, as the separation of the images formed is of order microarcseconds, and hence what is observed is an increase in brightness rather than multiple images. 
The typical timescale for microlensing of a star in the Magellanic Clouds (MC) by a CO in the Milky Way (MW) halo is approximately
\begin{equation}
 t \sim 1 \, {\rm yr} \left( \frac{M}{M_{\odot}} \right)^{1/2} \,.
\end{equation}
Therefore to probe multi-solar mass CO a long-duration survey is required, while to probe planetary mass CO a high-cadence survey is needed.

Multiple microlensing surveys have regularly monitored tens of millions of stars in the MC over long time periods. In the late 1990s, the MACHO collaboration observed more microlensing events than expected from known stellar populations, consistent with tens of per cent of the MW halo being composed of roughly Solar mass CO~\cite{MACHO:1995udp,MACHO:2000qbb}. However this result has not been confirmed by subsequent surveys, with the EROS collaboration excluding COs with mass $10^{-6} M_{\odot} \lesssim M \lesssim 10 M_{\odot}$ making up all of the MW halo~\cite{EROS-2:2006ryy}. More recently, using 20 years of data, the OGLE collaboration has excluded $f_{\rm PBH}=1$ up to $M \sim 10^{4} M_{\odot}$~\cite{Mroz:2024mse}. OGLE have also carried out a high-cadence survey of the MC and have constrained planetary mass PBHs ($10^{-8} M_{\odot} \lesssim M \lesssim 0.01 M_{\odot}$) to make up less than 1\% of the MW halo~\cite{Mroz:2024wia}. A high-cadence microlensing survey of M31 (Andromeda) has excluded $f_{\rm PBH}=1$ down to $M \sim 10^{-11} M_{\odot}$~\cite{Niikura:2017zjd}. PBH with mass below $M \lesssim 10^{-12} M_{\odot}$ can not be probed by optical microlensing observations as the microlensing magnification is reduced by finite source effects and wave optics (the wavelength of light is similar to the Schwarzschild radius of the PBH), see Ref.~\cite{Sugiyama:2019dgt} and references therein.



\subsection{Other direct constraints}
\label{sec-other}

As well as the stellar microlensing constraints, there are also gravitational lensing constraints on the abundance of CO (including PBHs) from supermagnified stars, flux ratios of multiply-lensed quasars and the magnification distribution of type 1a supernovae. If all of the DM were in the form of multi-Solar mass PBHs then the merger rate of PBH binaries is expected to be larger than that observed by LIGO-Virgo, so these gravitational wave observations in fact constrain $f_{\rm PBH}$. Multi-Solar mass PBHs can also be constrained by the consequences of their accretion, specifically the effect of the subsequent radiation on the recombination history of the Universe and hence the CMB, and also present-day X-ray or radio emission. Massive PBHs can be constrained via their dynamical effects on stars, in wide binaries and dwarf galaxies. For more detailed discussion of these constraints, including references to the original papers, see e.g.~Refs.~\cite{Green:2020jor,Escriva:2022duf,PBHbook}.

\subsection{Indirect constraints on large density perturbations}
\label{sec-indirect}

The direct constraints on the PBH abundance that we have discussed so far in this Section apply to all PBHs, whatever their formation mechanism. In this subsection, we briefly overview constraints on the amplitude of density perturbations, which indirectly constrain the abundance of PBHs formed from the collapse of large density perturbations. Large scalar (i.e.~density) perturbations generate tensor perturbations at second order~\cite{Ananda:2006af}, known as `scalar induced gravitational waves'. Constraints on the energy density of stochastic gravitational waves therefore constrain the amplitude of density perturbations and hence the abundance of PBHs formed via the collapse of large density perturbations~\cite{Saito:2008jc}. Observations from pulsar timing arrays such as NANOgrav potentially place tight constraints on PBHs formed via this mechanism for $10^{-3} 
M_{\odot} \lesssim M \lesssim 1 M_{\odot}$, for an overview see Ref.~\cite{Yuan:2021qgz}. Spectral distortions of the CMB constrain the primordial curvature perturbation power spectrum ~\cite{Carr:1993aq,Kohri:2014lza}, ${\cal P}_{\cal R} < 10^{-2}$ on scales $1 \, {\rm Mpc}^{-1} \lesssim k \lesssim 10^{5} \, {\rm Mpc}^{-1}$ which excludes $f_{\rm PBH} =1$ for $10^{3} M_{\odot} \lesssim M \lesssim 10^{12} M_{\odot} $ for PBHs formed from the collapse of large density perturbations.

\subsection{Probing asteroid mass Primordial Black Holes}
\label{sec-asteroid}
As shown in Fig.~\ref{fig:constraints}, PBHs in the so-called `asteroid mass window' ($10^{17} \, {\rm g} \lesssim M \lesssim 10^{22} \, {\rm g}$) are currently allowed to make up all of the DM ($f_{\rm PBH}=1$).  The $M^{-2}$ dependence of the PBH mass loss rate means that future observations can only increase the largest mass for which $f_{\rm PBH}=1$ is excluded by evaporation constraints by a factor of a few, while finite source and wave optics effects prevent optical microlensing observations probing PBH masses below $\sim 10^{22} \, {\rm g}$.
New techniques are required to probe all of the asteroid mass window, for instance, gamma-ray burst lensing parallax~\cite{Nemiroff:1995ak}, microlensing of X-ray pulsars~\cite{Bai:2018bej}, the consequences of PBH encounters of stars (see e.g. Ref.~\cite{Montero-Camacho:2019jte} and references therein) and the effects of PBHs on the orbits of planets~\cite{Tran:2023jci} and satellites within the Solar System~\cite{Cuadrat-Grzybowski:2024uph}. For more detailed overviews see e.g.~Refs.~\cite{Green:2020jor,Escriva:2022duf}.

\section{Conclusions}
\label{sec-conclusions}

In Sec.~\ref{sec-form} we studied the formation of PBHs, focusing on the most popular mechanism, the collapse during radiation domination of large density fluctuations generated by inflation. In this case, the amplitude of the perturbations has to be several orders of magnitude larger than the value measured on cosmological scales. This can be achieved in a subset of inflation models with, for instance, a feature in the potential. The amplitude of the perturbations also has to be fine-tuned to produce PBHs which make up a non-negligible fraction of the dark matter. When calculating the abundance (and mass function) of PBHs it is usually assumed that the distribution of the sizes of the density perturbations is gaussian. However, this is likely to not be the case for models which produce large, PBH-forming, fluctuations. Therefore an accurate calculation of the probability distribution of the density perturbations is a major open problem for PBH formation.

"What is dark matter?" is a question that will be answered by observations and experiments rather than theoretical calculations or arguments. Section \ref{sec-obs} explored the wide range of observations via which the abundance of PBHs can be probed. While there are some observations that can potentially be explained by PBHs (see Ref.~\cite{Carr:2023tpt} for a review), taken at face value, current constraints exclude PBHs from making up all of the DM, apart from in the `asteroid mass window' ($10^{17} \, {\rm g} \lesssim M \lesssim 10^{22} \, {\rm g}$). Whether or not the assumptions made in calculating these constraints, for instance about the spatial distribution of PBHs, are reliable is a key question. Using new data, and coming up with new ideas, to probe the asteroid mass window will also be a major focus in the future. See Ref.~\cite{Riotto:2024ayo} for a discussion of the future of PBH studies as of 2024.

\begin{ack}[Acknowledgments]%
AMG is supported by STFC grant ST/P000703/1. 
\end{ack}

\seealso{\begin{itemize}
\item{`Primordial Black Holes as a dark matter candidate', Green \& Kavanagh  \cite{Green:2020jor} \\
A concise 2020 review covering PBH formation and abundance constraints.}
\item{`Primordial black holes as dark matter candidates', Carr \& K\"uhnel \cite{Carr:2021bzv} \\
Lecture notes from the 2021 Les Houches summer school on Dark Matter.}
\item{`Primordial black holes', Escriv\`a, K\"uhnel \& Tada \cite{Escriva:2022duf} \\
A detailed 2022 review.}
\item {`Lecture notes on inflation and primordial black holes', Byrnes \& Cole \cite{Byrnes:2021jka} \\
Based on a short lecture course held at the Galileo Galilee Institute in 2021.}
\item{`Observational evidence for primordial black holes: a positivist perspective', Carr, Clesse, Garcia-Bellido, Hawkins \& K\"uhnel \cite{Carr:2023tpt} \\
A 2024 review focusing on potential evidence for PBHs.}
\item{`Primordial Black Holes', Editors: Byrnes, Franciolini, Harada, Pani \& Sasaki \cite{PBHbook}  \\
A 2024 book providing a very comprehensive overview.}
\item{`Primordial black hole numbers: standard formulas and chart', Tomberg \cite{Tomberg:2024chk}.\\
A short note containing derivations of accurate expressions for the mass and abundance of PBHs and useful charts relating key quantities.}
\end{itemize}}

\bibliographystyle{Numbered-Style} 
\bibliography{reference}

\end{document}